\documentclass{cernyrep}
\usepackage{varwidth}
\usepackage{xcolor}

\begin{document}

\title{Machine Protection}
\author{R.~Schmidt}
\institute{CERN, Geneva, Switzerland}
\maketitle

\begin{abstract}
The protection of accelerator equipment is as old as accelerator technology and was for many years related to high-power equipment. Examples are the protection of powering equipment from overheating (magnets, power converters, high-current cables), of superconducting magnets from damage after a quench and of klystrons. The protection of equipment from beam accidents is more recent. It is related to the increasing beam power of high-power proton accelerators such as ISIS, SNS, ESS and the PSI cyclotron, to the emission of synchrotron light by electron--positron accelerators and FELs, and to the increase of energy stored in the beam (in particular for hadron colliders such as LHC). Designing a machine protection system requires an excellent understanding of accelerator physics and operation to anticipate possible failures that could lead to damage. Machine protection includes beam and equipment monitoring, a system to safely stop beam operation (e.g.\ dumping the beam or stopping the beam at low energy) and an interlock system providing the glue between these systems. The most recent accelerator, the LHC, will operate with about
$3 \times 10^{14}$ protons per beam, corresponding to an energy stored in each beam of 360~MJ. This energy can cause massive damage to accelerator equipment in case of uncontrolled beam loss, and a single accident damaging vital parts of the accelerator could interrupt operation for years. This article provides an overview of the requirements for protection of accelerator equipment and introduces the various protection systems. Examples are mainly from LHC, SNS and ESS.
\end{abstract}

\def\labelenumi{\roman{enumi})}

\section{Introduction}

Accelerators, like all other technical systems, must respect some general principles with respect to safety and protection:
\begin{enumerate}
  \item protect the people (e.g.\ following legal requirements);
  \item protect the environment (e.g.\ following legal requirements);
  \item protect the equipment (the investment).
\end{enumerate}

The term 'equipment' includes all systems of the accelerator that must be protected during all phases of operation, with the accelerator operating with or without beam.

In this presentation the protection of equipment from damage or unacceptable activation caused by beam losses is discussed. First, some principles for the protection of accelerators are presented and then examples for machine protection from SNS \cite{Sibley2003}, ESS \cite{Peggs2005} and LHC \cite{Schmidt2006} are given.

Protection is not only relevant during operation with a beam. Without a beam, protection of high-power equipment must be considered. This is not the main subject of the article; however, a few examples are given as follows.
\begin{enumerate}
\item Superconducting magnets store a large amount of energy and there is the risk of a quench, either induced by the beam or due to other reasons, e.g.\ training quenches. The magnets need to be designed and protected so that they are not damaged after a quench. The incident at the LHC on 19 September 2008 demonstrated that the risk of operation of a superconducting magnet system is high \cite{Bajko2009}.
\item Some equipment (normal-conducting magnets, high-current cables, high-current power converters) requires air or water cooling. In case of a failure in the cooling system the equipment needs to be switched off.
\item In radiofrequency (RF) systems there is the risk of arcing due to high voltages. Arc detectors detect arcs and switch off the high-voltage source to avoid any damage.
\end{enumerate}

Although there are quite a number of publications on issues related to machine protection, there are only a few summary papers, e.g.\ \cite{Sibley2003}.

\section{Definition of risk}
Risks come from energy stored in a system (measured in joules) and power when operating a system (measured in watts). The energy and power flow needs to be controlled. An uncontrolled release of the energy, or an uncontrolled power flow, can lead to unwanted consequences:
\begin{enumerate}
  \item damage to equipment and loss of time for operation;
  \item for particle beams, activation of equipment.
\end{enumerate}

This is true for all systems, in particular for complex systems such as accelerators.

A `hazard' is a situation that poses a level of threat to the accelerator.
Hazards are dormant or potential, with only a theoretical risk of damage. Once a hazard becomes `active' it becomes an incident or accident. Consequences and probability of an incident interact together to create a risk that can be quantified:
\begin{equation}
    \mathrm{Risk} = \mathrm{Consequences} \times \mathrm{Probability}.
\end{equation}

Related to accelerators, the consequences and the probability of an uncontrolled beam loss need to be estimated to get an idea about the risk. Machine protection systems prevent damage to equipment after a failure. The higher the risk, the more machine protection becomes important. Machine protection needs to be considered during design, construction and operation of the accelerator.

If a specific failure is considered, the consequences of the failure can be estimated, in terms of damage to equipment (repair requiring investment, e.g.\ in money), in downtime of the accelerator (e.g.\ in days) and in radiation dose to personnel accessing equipment (e.g.\ in mSv).

In the estimation of downtime of the accelerator for repairs the availability of spare parts needs to be considered. If the accelerator was operating with beam, radioactive activation of material must be taken into account. It may be necessary to wait for cool-down of irradiated components to reduce the dose before accessing the equipment.

The second factor entering into the risk is the probability of such a failure happening (e.g.\ measured in number of failures per year).

For beam operation, a list of all possible failures that could lead to beam loss in equipment should be considered. This is not obvious, since there is a nearly infinite number of mechanisms for losing the beam. However, the most likely failure modes and in particular the worst-case failures and their probabilities must be considered.

\section{Challenges in the protection of LHC and ESS}
Many accelerators operate with high beam intensity and/or energy. For synchrotrons and storage rings, the energy stored in the beam increased with time (from ISR to LHC). For linear accelerators and fast-cycling machines, the beam power increases. Two examples illustrate machine protection challenges for a circular collider (LHC) and a high-intensity proton linac (ESS).

For the future another challenge is the  emittance that becomes smaller, down to a beam size of a nanometer. This is important today, and even more relevant for future projects such as ILC, with increased beam power or energy density (measured in W/mm$^2$ or J/mm$^2$) and increasingly complex machines.

\begin{figure}
  \centering
  \includegraphics[width=.75\linewidth]{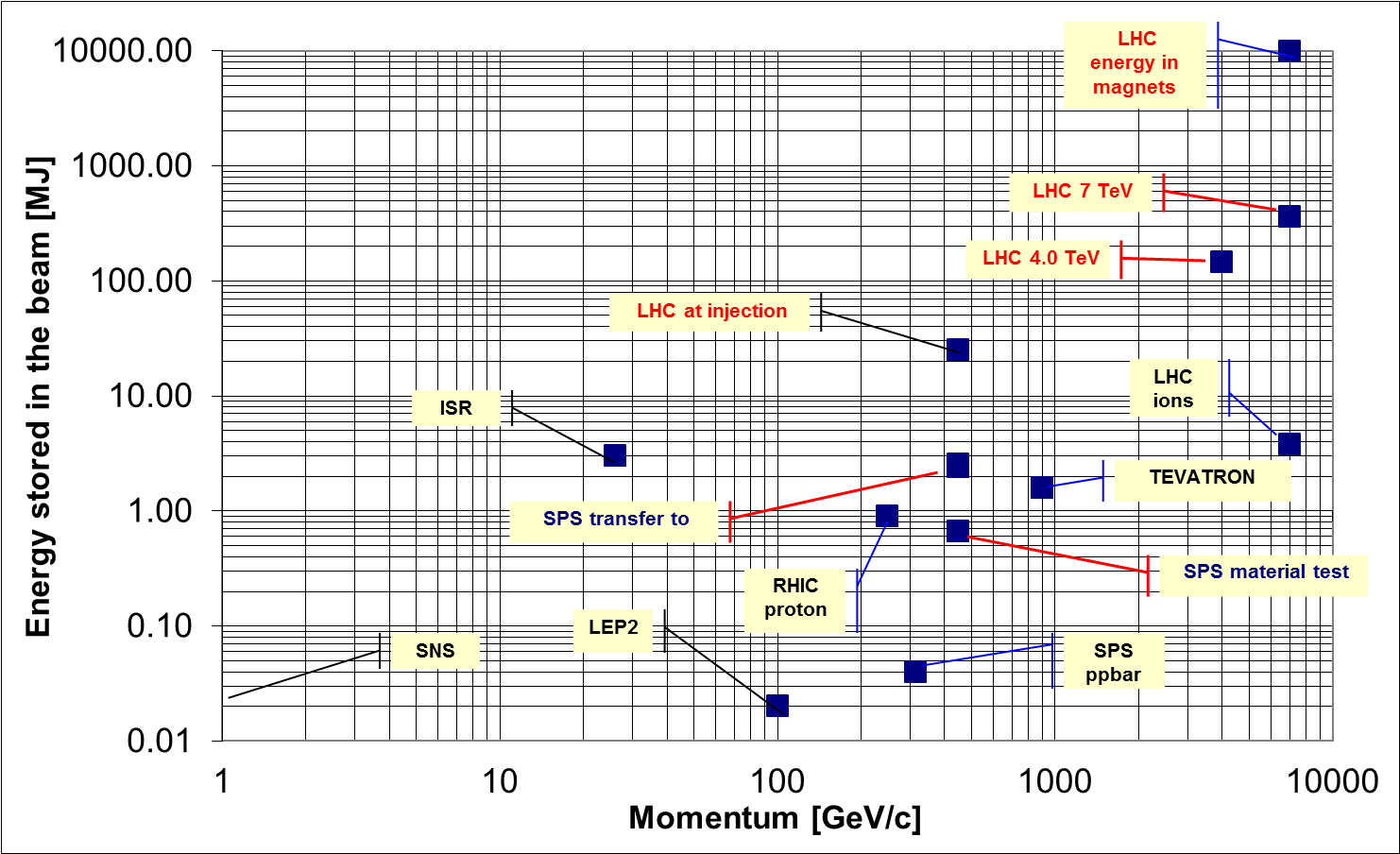}
  \caption{Energy stored in the beams for different accelerators (based on a figure by R.~Assmann)}
  \label{fig-stored-beam-energy}
\end{figure}

\subsection{LHC collider}
The LHC is designed to operate at a momentum of 7~TeV/$c$ with 2808 bunches, each bunch with a nominal intensity of $1.15 \times 10^{11}$ protons. Machine protection is required during all phases of operation since the LHC is the first accelerator with the intensity of the injected beam already far above the threshold for damage. The energy stored in the nominal LHC beam of 362~MJ corresponds to the energy of a 200~m long fast train at 155 km per hour and to the energy stored in 90~kg of TNT. Surprisingly, this is the same as the energy stored in 15~kg of chocolate; it matters most how easily and fast the energy is released. The energy in an accelerator beam can be released in some 10~$\mu$s.

A Livingston-type plot (\Fref{fig-stored-beam-energy}) shows the energy stored in the beam as a function of particle momentum. At 7~TeV/$c$, fast beam loss with an intensity of about 5\% of a single `nominal bunch' could damage equipment (e.g.\ superconducting coils). The only component that can stand a loss of the full beam is the beam dump block. All other components would be severely damaged. The LHC beams must \textit{always} be extracted into the beam dump blocks at the end of a fill as well as in case of a failure.

\begin{figure}[t]
\centering
\includegraphics[width=.6\linewidth]{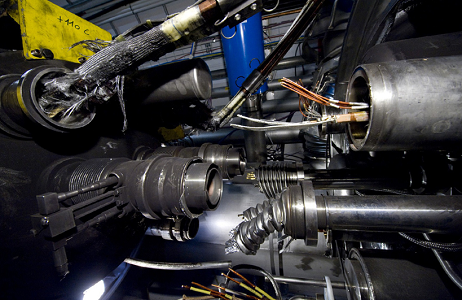}
\caption{Damage from an electrical arc after the LHC accident in 2008}
\label{LHC-accident-2008-arcing}
\end{figure}

\begin{figure}[t]
\centering
\includegraphics[width=.42\linewidth]{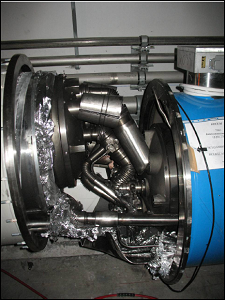}
\caption{Damage from the overpressure in the cryostat after the LHC accident in 2008}
\label{LHC-accident-2008-overpressure}
\end{figure}

\begin{figure}[t]
\centering
\includegraphics[width=.6\linewidth]{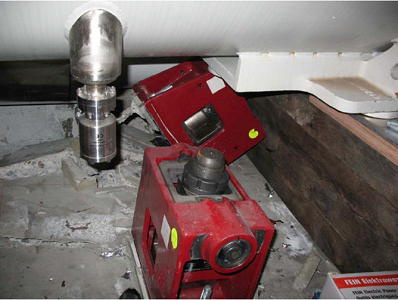}
\caption{Displacement of a magnet after the LHC accident in 2008}
\label{LHC-accident-2008-displacement}
\end{figure}

During the first phase of operation between 2009 and 2013 the momentum was limited to 4~TeV/$c$ and the maximum stored beam energy was up to about 140~MJ. This was the consequence of the 2008 LHC accident that happened during test runs without beam. A magnet interconnect was defective and the circuit opened. An electrical arc provoked a helium pressure wave damaging about 600~m of the LHC and polluting the beam vacuum over more than 2~km (Fig. \ref{LHC-accident-2008-arcing}). An overpressure from the expansion of liquid helium damaged the structure (see Figs. \ref{LHC-accident-2008-overpressure} and \ref{LHC-accident-2008-displacement}). A total of 53 magnets had to be repaired.

Protection during the injection process is mandatory. The energy stored in the LHC beam at injection is about one order of magnitude higher than the stored energy in the beam for other accelerators.

\begin{figure}[t]
\centering
\includegraphics[width=.48\linewidth]{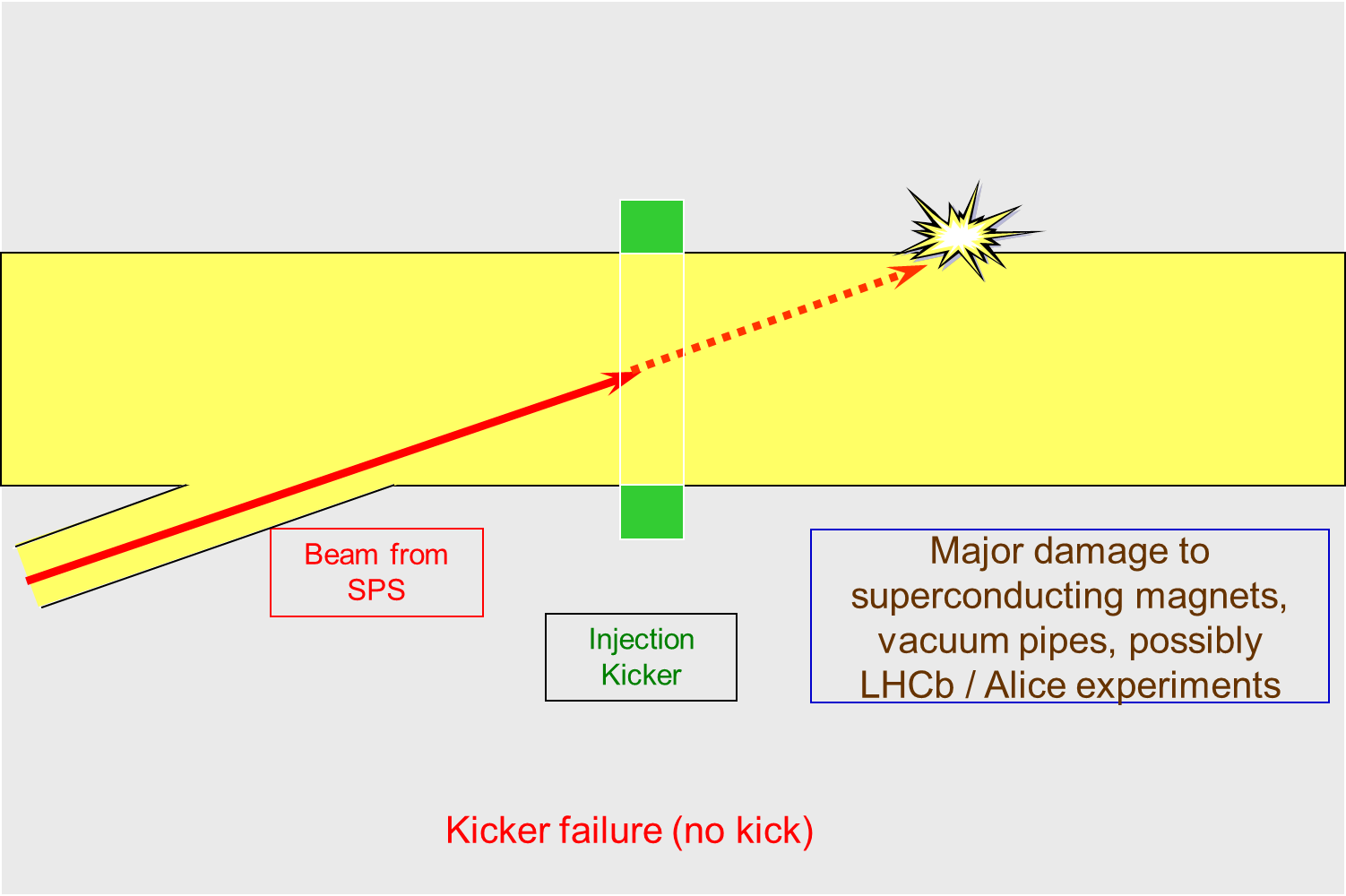}
\includegraphics[width=.48\linewidth]{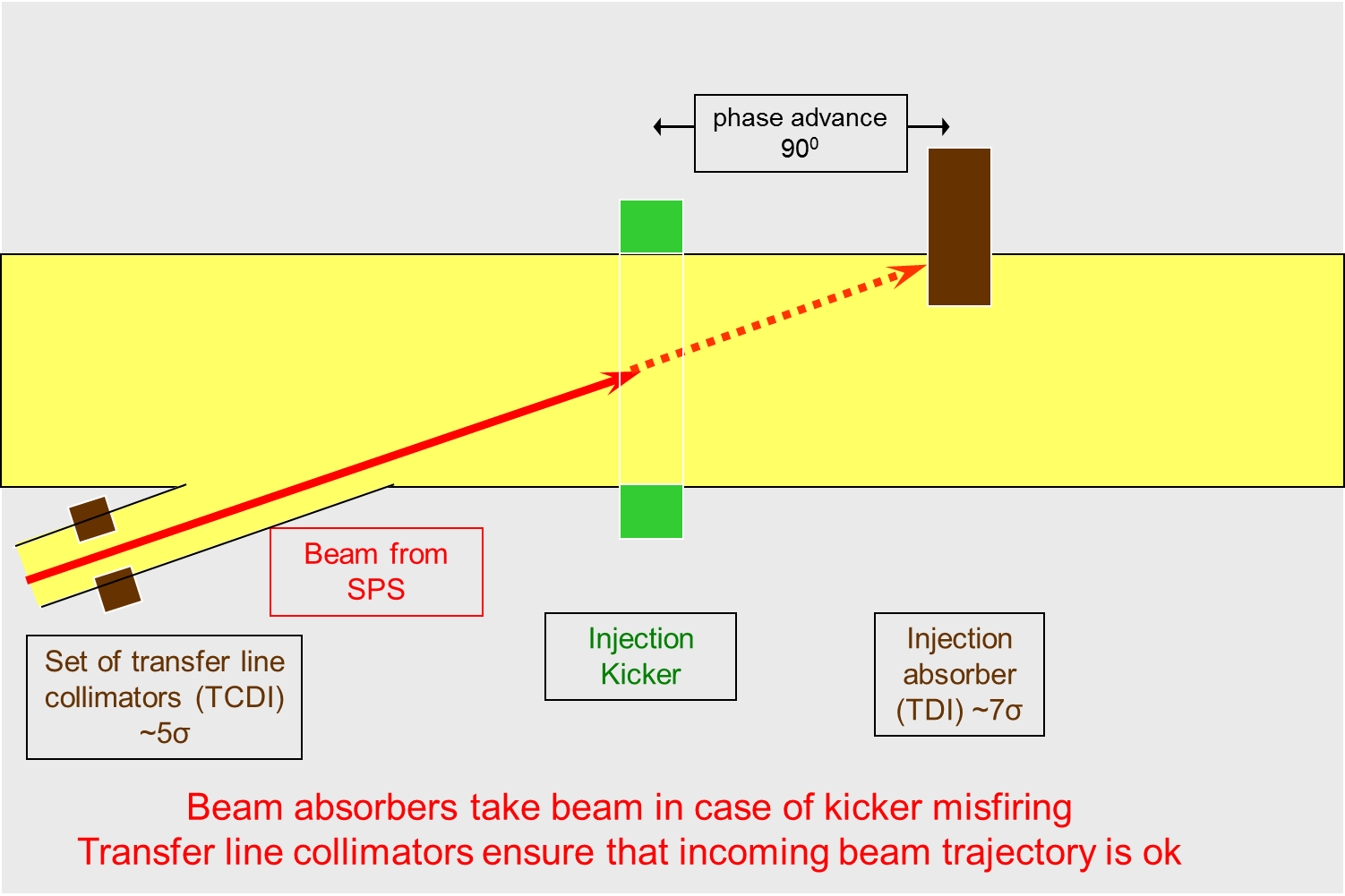}
\caption{LHC injection failure. Left: the injection kicker does not deflect the incoming beam. The beam would hit vacuum chamber and equipment. Right: the injection kicker does not deflect the incoming beam. The injection absorber is correctly positioned and the beam hits the absorber.}
\label{LHC-injection-failure}
\end{figure}

An example of a critical failure for the LHC at injection is a failure of the injection kicker. If the kicker does not fire or fires at the wrong time, the incoming beam will not be deflected or the circulating beam will be deflected (see Fig. \ref{LHC-injection-failure}). In both cases, the beam would hit the vacuum chamber and damage equipment.

During the injection process injection collimators are positioned in the vacuum chamber to capture the mis-kicked beam. This type of failure, e.g.\ a wrong kick of the injection kicker, happened already several times. Since the injection absorber was always at the correct position no damage was observed; however, due to grazing beam incidence superconducting magnets downstream of the injection region quenched.

\subsection{ESS spallation source}
\begin{figure}[t]
\centering
\includegraphics[width=1.0\linewidth]{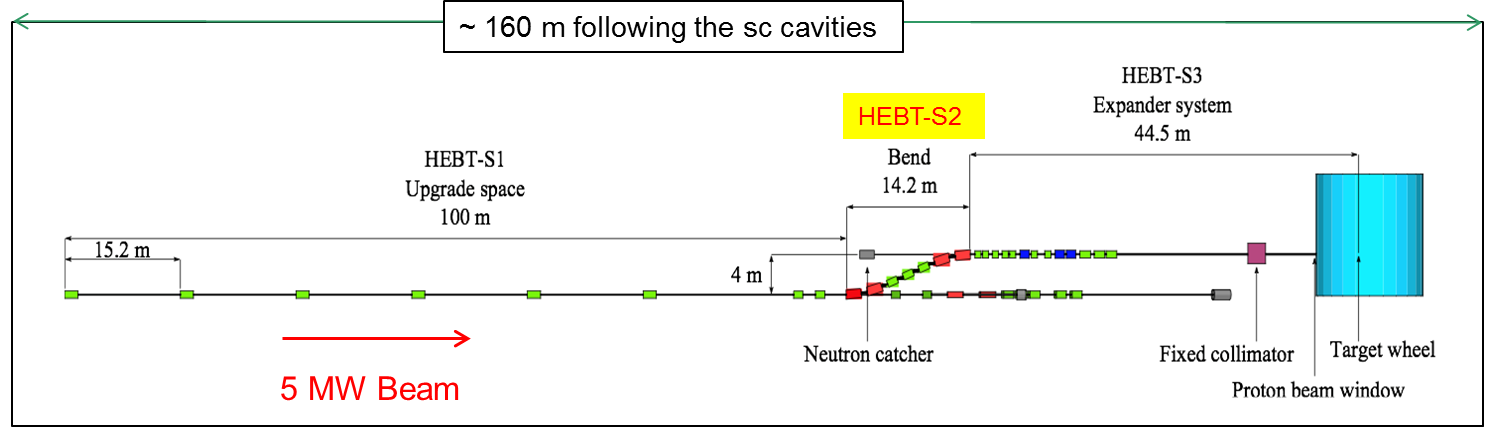}
\caption{High-energy part of the ESS accelerator. The beam is bent from the accelerator plane upwards to the plane with the high-power target. It is also possible to let the beam go straight to a commissioning beam dump.}
\label{ESS-bend}
\end{figure}

The ESS spallation source being built at Lund, Sweden is designed to accelerate a proton beam with an average power of 5~MW and to direct it onto a target. Operation of the ESS will be  at a frequency of 14~Hz, with a pulse length of 2.86~ms with a peak power of 125~MW. The layout of the ESS accelerator is shown in Fig.~\ref{ESS-layout}. We now give an example of a critical failure at ESS (see Fig. \ref{ESS-bend}):

\begin{figure}[t]
\centering
\includegraphics[width=1.0\linewidth]{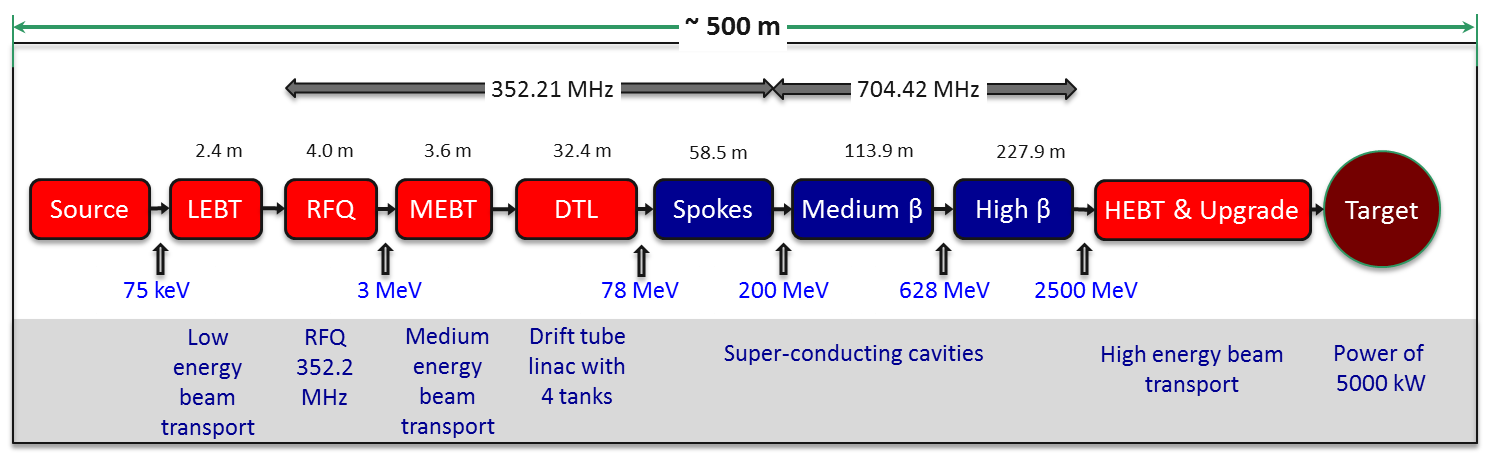}
\caption{Layout of the ESS accelerator: the source, low-energy beam transport and RFQ are followed by the medium-energy beam transport. The protons are accelerated by a normal-conducting linac, followed by three sections of superconducting cavities. In the high-energy beam transport line the protons are transported to the target.}
\label{ESS-layout}
\end{figure}

\begin{enumerate}
  \item the bending magnet in the junction between HEBT-S2 and HEBT-S3 deflects the beam to bring the protons into the target plane;
  \item assume that the power supply for the bending magnets in HEBT-S2 fails and the magnets stop deflecting the beam;
  \item the probability for such failure in one year is about 0.1 (assuming a mean time between failures (MTBF) for power supplies of 700,000 h = 10 years);
  \item the beam is not deflected and hits the vacuum chamber. The consequences would be damage to magnets and vacuum pipes and possibly a pollution of superconducting cavities upstream.
\end{enumerate}

If a failure is detected the beam must be stopped at the source or in the low energy part of the accelerator. 

\section{Beam losses and consequences}

\subsection{Energy deposition from high-energy particles}
Particle losses in a material lead to ionization and particle cascades. Energy deposition starts when a charged particle enters material. The maximum energy deposition can be deep in the material at the maximum of the hadron or electromagnetic shower depending on the particle momentum. The energy deposition leads to a temperature increase in the material that can be vaporized, melted, deformed or lose its mechanical properties, depending on the material and the beam impact.

There is already some risk of damage to sensitive equipment for an energy deposition of some 10~kJ (beam impact for a short time, say a maximum of a few milliseconds).
The risk of damage to equipment  for some megajoules is very large.

Equipment becomes activated due to beam losses. It is considered to be acceptable if beam losses do not exceed, say, 1~W/m (assuming proton beams with high energy, say above some 100~MeV/$c$). Another principle is `ALARA': exposure of personnel to radiation should be `as low as reasonably achievable'. If a further reduction of the beam losses below 1~W/m is reasonably possible, this is recommended in order to minimize exposure of service personnel. Radioactive activation of material is mainly an issue for hadron accelerators; it is less problematic for electron--positron machines.

For accelerators with superconducting magnets there is a specific problem: even with beam loss much below the damage threshold, superconducting magnets could quench. In case of a quench, beam operation is interrupted for some time (10 min up to many hours) leading to downtime. In order to avoid beam-induced quenches, beam losses are monitored and the beam is dumped if a predefined threshold is exceeded before a magnet quenches, reducing the downtime since the time to recover from a quench is avoided. The damage threshold is far above the quench threshold; therefore, this strategy also protects magnets from beam-induced damage.

\begin{figure}[t]
\centering
\includegraphics[width=.65\linewidth]{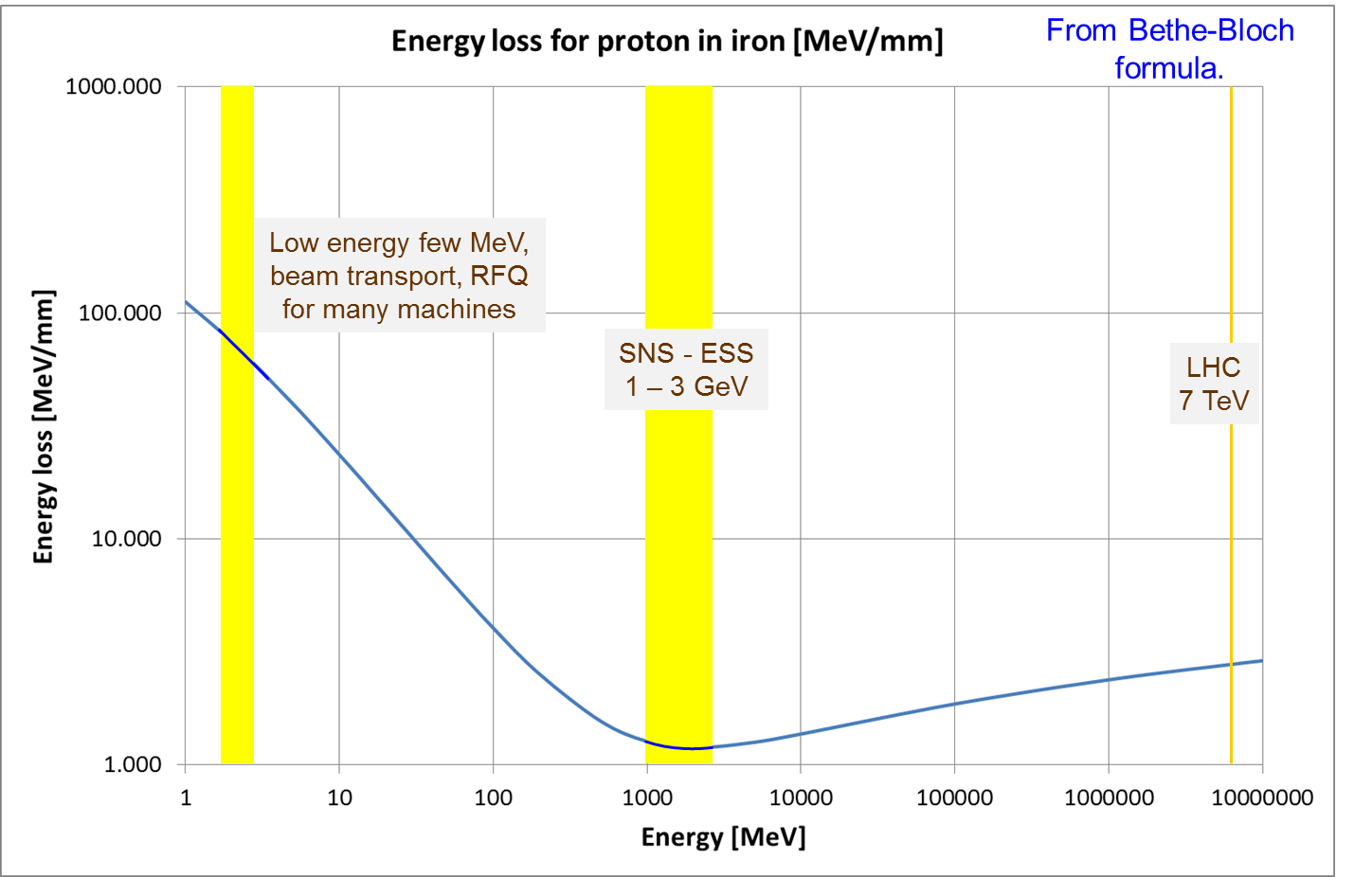}
\caption{Energy loss for protons when entering into an iron block, calculated by the Bethe--Bloch equation. Effects of the hadron shower are not included.}
\label{Energy-Loss-Iron-BB}
\end{figure}

To get an idea of the damage potential, the energy loss by ionization at the surface of the target can be estimated using the Bethe--Bloch equation, ignoring hadron showers (see Fig.~\ref{Energy-Loss-Iron-BB}). It is interesting to observe that the energy loss for low-energy particles is very high. For a proton at 7~TeV/$c$ the energy loss is dominated by hadron showers and the maximum energy deposition is deep in the material.

There is no straightforward expression for the energy deposition of high-energy particles, since this depends on the particle type, the momentum, the beam parameters and the material parameters (atomic number, density, specific heat). Programs such as FLUKA \cite{Fasso2003}, MARS \cite{MARS} or GEANT4 \cite{GEANT} are being used for the calculation of energy deposition (and subsequent temperature increase) as well for as the activation of the material that is exposed.

Relevant parameters to be considered for heating and possible damage to material are the momentum of the particle, the particle type, the energy stored in the beam, the beam power, the beam size, the beam power/energy density (MJ/mm$^{2}$, MW/mm$^{2}$), the time structure of the beam and cooling conditions.

In order to estimate the order of magnitude for possible damage:
\begin{enumerate}
  \item one MJ can heat and melt about 1.5~kg of copper;
  \item one MJ corresponds to the energy stored in about 0.25~kg of
  TNT \cite{wiki_1};
  \item one MW during one second corresponds to one MJ.
\end{enumerate}

The following example illustrates the calculation of the consequences for beam losses.
\begin{enumerate}
  \item A proton beam travels through a thin window of thickness $d$.
  \item Assume a beam area of 4 $\sigma_x \times \sigma_y$, with $\sigma_x$ and $\sigma_y$ the root mean square (r.m.s.) beam sizes assuming Gaussian beams.
  \item Assume a homogeneous beam distribution.
\end{enumerate}

The energy deposition can be calculated; mass and specific heat are known.
The temperature can be calculated (rather good approximation), assuming a fast loss and no cooling.

With low-energy protons (3 MeV/$c$) and a beam size of $\sigma_x = \sigma_y = 1$~mm, the following parameters are calculated:
\begin{enumerate}
  \item iron specific heat = 440 J/(kg $\times$ K);
  \item iron specific weight = 7860 kg/m$^3$;
  \item d$E$/d$x$ = 56.7 MeV/mm;
  \item $N_{\rm p} = 1.16 \times 10^{12}$.
\end{enumerate}
The temperature increase for these parameters is 763~K.

\begin{figure}[t]
\centering
\includegraphics[width=.4\linewidth]{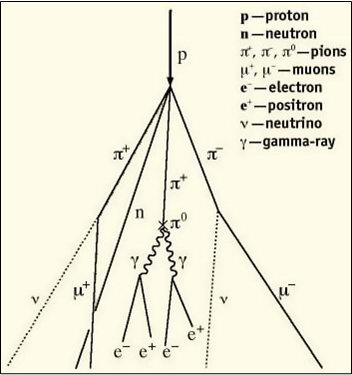}
\vspace{0.5cm}
\caption{Illustration of a hadronic shower}
\label{Hadron-Shower}
\end{figure}

For high-energy protons the energy deposition by hadronic showers (see Fig. \ref{Hadron-Shower}) dominates:
\begin{enumerate}
    \item pions are created when the proton travels through matter;
    \item the decay of pions creates electromagnetic showers;
    \item there is an exponential increase in number of created particles;
    \item the final energy deposition is to a large extent due to the large number of electromagnetic particles;
    \item the energy deposition scales roughly with total energy of incident particles;
    \item the maximum of the energy deposition can be deep in the material;
    \item energy deposition is a function of the particle type, its momentum and parameters of the material (atomic number, density, specific heat).
\end{enumerate}

There is no straightforward expression to calculate the energy deposition by high-energy particles. Calculation needs to be done by codes, such as FLUKA, GEANT or MARS.

A simple approximation for the temperature increase in material for a 7~TeV/$c$ proton beam impact is given in the following example: for copper, the maximum longitudinal energy deposition for a single 7~TeV/$c$ proton at about 25~cm inside the material is $E_{\rm dep}=1.5 \times 10^{-5}$~J/kg (calculation with FLUKA).

The energy required to heat and melt copper is $E=6.3 \times 10^5$~J/kg. Assuming a pencil beam, the number of particles required to damage (melt) copper is of the order of $10^{10}$. For graphite, the number of particles needed to cause damage is about one order of magnitude larger.

More refined and complete calculations can be made to determine real-world scenarios on a case-by-case basis, where the distribution of the impacting particles and the details of the material and geometry are important.  After the calculation of the temperature increase, the response of the material to beam impact needs to be addressed (deformation, melting etc). Mechanical codes such as ANSYS and  hydrodynamic codes such as BIG2 and others can be used.

Beams at very low energy have limited power; however, the energy deposition is very high and can lead to (limited) damage in the case of a beam impact at the initial stage of an accelerator, after the source, in the low-energy beam transport and in the radiofrequency quadrupole (RFQ). This might lead to a long downtime, depending on availability of spares.

Beams at very high energy can have a tremendous damage potential for the LHC; damage to metals is expected for about $10^{10}$ protons. One LHC bunch has about $1.5\times10^{11}$ protons, in total up to 2808 bunches. In case of catastrophic beam loss, the LHC could be possibly damaged beyond repair. The penetration of 2808 bunches (e.g.\ after a kicker failure) into a metal block such as a magnet has been estimated to be about 20 to 30~m (hydrodynamic beam tunnelling) \cite{Tahir2005}.

\subsection{Examples for damage from beam losses}
The design of LHC protection elements is based on detailed energy deposition simulations and an assumption for the damage levels. A dedicated experiment was carried out to cross-check the validity of this approach by trying to damage material in a controlled way with beam~\cite{Kain2005a}. The impact of a 450~GeV/$c$ beam extracted from the SPS on a specially designed high-$Z$ target was simulated for a simple geometry comprising several typical materials used for LHC equipment. The beam intensities for the test were chosen to exceed the damage limits of parts of the target, between $2 \times 10^{12}$ and $8 \times 10^{12}$ protons. The transverse r.m.s.\ beam dimensions were about 1~mm.

\begin{figure}
  \centering
  \includegraphics[width=.6\linewidth]{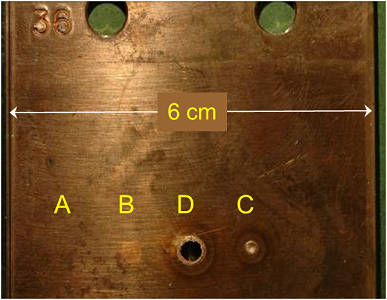}
  \caption{Damage to a copper plate by 450 GeV/$c$ proton beams with different intensities ($2\times 10^{12}, 4\times 10^{12}, 6\times 10^{12} $ and $ 8\times 10^{12}$).}
  \label{DamageExperiment}
\end{figure}

The geometry of the target was modelled in FLUKA and the target heating was estimated. The temperature rise was obtained from the energy deposition using the temperature-dependent heat capacity for each material. The results of the controlled damage test show reasonable agreement with the simulations. Zinc, copper and INCONEL plates were damaged at the predicted locations within the error of the simulation, see Fig.~\ref{DamageExperiment}. The transverse extent of the damaged area on the zinc and copper plates was predicted to within 30\%. The outcome of the experiment gives confidence that beam-induced damage limits for simple geometries can be adequately predicted with simulations.

An example for involuntarily beam-induced accidents is the damage to a vacuum chamber in the SPS extraction line. In 2004, during extraction tests with LHC beam parameters with 450~GeV/$c$ protons, a beam with an energy of 2~MJ was deflected with grazing incidence into a vacuum chamber after the failure of a septum magnet. The chamber was cut along a length of 25~cm, with a groove of a length of 70~cm. Condensed drops of steel are visible on the opposite side of the vacuum chamber. Vacuum chamber and quadrupole magnet needed to be replaced.

\begin{figure}
  \centering
  \includegraphics[width=.6\linewidth]{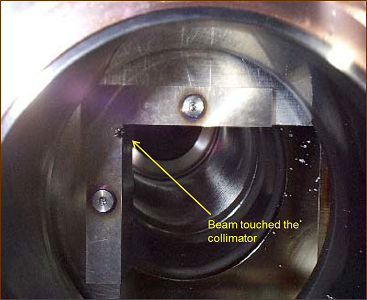}
  \caption{Damage to a collimator at Fermilab for 900 GeV/$c$ proton beams, when a Roman pot moved into the beam and massively quenched magnets.}
  \label{Damage-Roman-Pot}
\end{figure}

Another example for beam-induced damage is an accident at the Tevatron, Fermilab. A so-called Roman pot (a device equipped with particle detectors that can be moved in the vacuum chamber) moved into the beam. Particle showers generated by the Roman pot quenched superconducting magnets. The beam moved by 0.005 mm/turn, and touched a collimator jaw surface after about 300 turns. The entire beam was lost, mostly on the collimator that was damaged (see Fig. \ref{Damage-Roman-Pot}).

Observation of HERA/DESY showed grooves on the surface of the tungsten collimator when opening the vacuum chamber. No impact was observed during operation.

Equipment that operates at high voltage can be very sensitive to beam losses. If the beam hit the surface of a high-voltage system (e.g.\ RFQ, kicker magnets, cavities) the surface quality might degrade. It is then not possible to operate at the same voltage and the probability for arcing is increased.

\begin{figure}
  \centering
  \includegraphics[width=.6\linewidth]{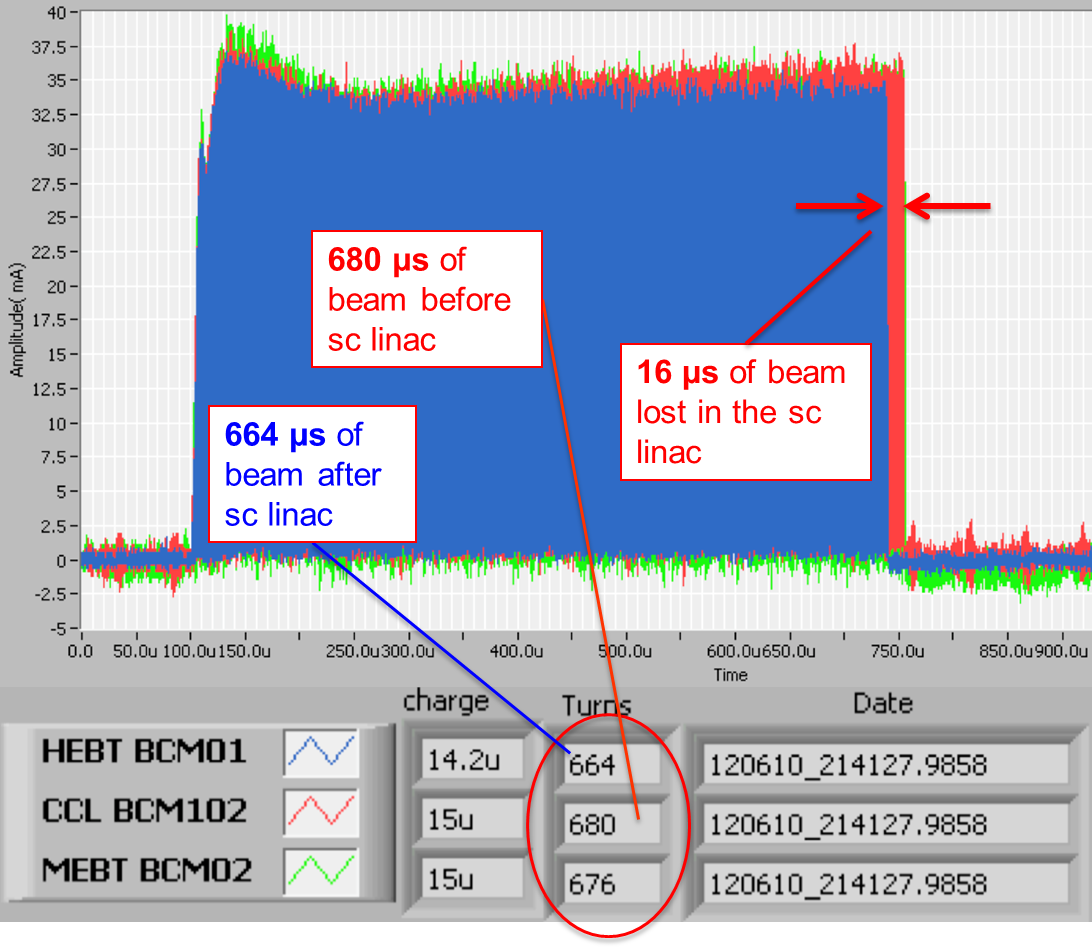}
  \caption{Beam losses in the SNS linac, with a 1 GeV/$c$ proton beam. These losses led to a degradation of superconducting cavities (figure kindly provided by M. Plum~\cite{Plum2013}).}
  \label{Damage-SNS}
\end{figure}

This was observed at SNS where errant beam losses led to a degradation of the performance of a superconducting cavity (see Fig. \ref{Damage-SNS} \cite{Blokland2013}). Beam current monitors (BCMs) measure current pulses at different locations along the linac, before the cavity section and after the cavity section (see the red and blue curves). The figure shows a beam loss during 16 $\mu$s; this corresponds to an energy of 30~J at the end of the Drift Tube Linac (DTL), 66~J at the end of the Coupled-Cavity Linac (CCL) and 350~J at the end of the superconducting linac.

Beam losses are likely to be caused by problems in the ion source, in the low-energy beam transfer or in the normal-conducting linac. After such errant beams, sometimes the cavity gradient needs to be lowered. Conditioning after warm-up helps in most cases, but in one case a cryo-module had to be exchanged. The energy of beam losses is about 100~J.
The damage mechanisms are not fully understood; it is assumed that some beam hitting the cavity desorbs gas or particulates (= small particles) creating an environment for arcing.

\section{Mechanisms for generating beam losses}
Continuous beam losses are inherent during the operation of accelerators and are taken into account during the design of the accelerator. Accidental beam losses are transient losses with time scales from ns to many seconds due to a multitude of failure mechanisms. The number of possible failures leading to accidental beam losses is (nearly) infinite.

`Machine protection' protects equipment from damage, activation and downtime in case of accidental beam losses. Machine protection includes a large variety of systems.

\subsection{Continuous beam losses}
Particles are lost due to a variety of reasons: beam--gas interaction, losses from collisions and losses of the beam halo. Limitation of beam losses is in order of 1~W/m to avoid activation and still allow hands-on maintenance for high hadron energy accelerators.

Several high-intensity hadron accelerators are operating with beams of a power of 1~MW and more (SNS, PSI). The are plans for accelerators of 5~MW (ESS), 10~MW and more. For 1~MW, a loss of 1\% corresponds to 10~kW, not to be lost along the beamline to avoid activation of material, heating, quenching of magnets etc. If a length of  200~m is assumed, the losses would correspond to 50~W/m, far above what is acceptable.

Another example is the LHC that operates with a stored beam with an energy of 360~MJ. A beam lifetime of 10~min corresponds to a beam loss of 500~kW that should not be lost in superconducting magnets. The strategy to keep beam losses at an acceptable level is as follows:
\begin{enumerate}
\item avoid beam losses as far as possible;
\item define the aperture by collimators;
\item capture continuous particle losses with collimators at specific locations.
\end{enumerate}

\begin{figure}
\centering
\includegraphics[width=.6\linewidth]{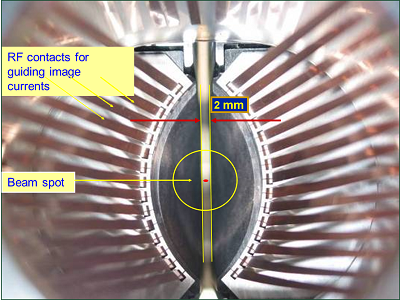}
\caption{Collimator at LHC}
\label{LHC-Collimator}
\end{figure}

A collimation system is a very efficient system to avoid too high beam losses in the accelerator (so-called beam cleaning). It can be very complex with (massive) material blocks close to the beam installed in an accelerator to capture halo particles. The collimation system at LHC reduces the losses by four orders of magnitude and also captures fast accidental beam losses. About 100 collimators are installed in LHC. Figure~\ref{LHC-Collimator} shows the view of one of the two-sided collimators, it is closed down to 2~mm when operating at 7~TeV/$c$. Collimators (or beam absorbers) are equally important to capture mis-steered beam.

\subsection{Accidental (transient) beam losses}
\subsubsection{Ultra-fast beam losses}
There are several mechanisms of failures that lead to a beam loss within a very short time, typically in the range of nanoseconds to microseconds.
\begin{enumerate}
\item Failures of kicker magnets (during injection, at extraction, for special kicker magnets for beam diagnostics).

\item Beam transfer via transfer lines between different accelerators and from the accelerator to a target station. The target could be for secondary particle production or the beam dump block. A common example for such a failure is a wrong setting of a magnet in the line that leads to a deflection of the beam into the vacuum chamber.

\item Too small beam size at a target station. Targets for high-intensity beams are designed for a certain beam size. If the beam size is much smaller, for example due to a wrong optics, the target or the window in front of the target could be damaged.
\end{enumerate}

\subsubsection{Very fast beam losses}
Mechanisms that lead to a very fast beam loss (typically in the range of some microseconds to milliseconds) are multiturn beam losses in a circular accelerator due to a large number of possible failures, mostly in the powering system of the main magnets, with a typical time constant of some 10 turns to many seconds. For linacs a trip of the RF can create beam losses within a very short time.

\subsubsection{Fast beam losses}
Fast beam loss (some 10 ms to seconds) can have many different origins: failures in the magnet powering system, vacuum valves that close unintentionally, trips of the RF acceleration system, beam instabilities, failures in the control system etc.

\subsubsection{Slow beam losses}
Slow beam losses (many seconds) can have many different origins: high vacuum pressure, failures in the powering system for corrector magnets, wrong parameter, RF etc. The main difference between slow and fast losses is that the operation crew could still be involved in the decision on how to continue operation (for example, to stop beam operation or to change a parameter that is not set correctly).

\subsection{Failure mechanisms and their probability}
\subsubsection{Failure of accelerator equipment}
Examples are power converter trips, magnet quenches, AC distribution failures (e.g.\ as a consequence of a thunderstorm), objects in the vacuum chamber, vacuum leaks, trip of the RF system, spontaneous firing of a kicker magnet etc.

\subsubsection{Failure of the control system}
Wrong parameter sent to equipment such as a wrong magnet current or a wrong magnet current function, trigger problem, failure in the timing system, failure in the feedback system, firing of a kicker magnet at the wrong time or with the wrong strength.

\subsubsection{Failure during operation}
Wrong manipulation of chromaticity, tune or orbit etc.

\subsubsection{Beam instability}
For example, due to too high beam or bunch current, with typical time constant of many milliseconds to seconds.

In order to calculate the risk related to a hazard, the probability of the occurrence and the damage potential need to be estimated. For the design of the protection system, the time constant of the failure as well as the time constant for beam loss due to the failure need to be known.

Experience from LHC (the most complex accelerator currently in operation): when the beams are colliding, the optimum length of a store is in the order of 10--15~h, then it is ended by operation with a beam dump. However, most fills (about 70\%) are ended by failures; the machine protection systems dump the beams. The MTBF is about 6~h. For other large accelerators (SNS, synchrotron light sources) the MTBF is between 20~h and up to several hundred hours.

\section{Strategy for machine protection}
\subsection{Principles for machine protection}
There are some principles for machine protection that need to be considered:
\begin{enumerate}
\item protect the machine;
\item protect the beam;
\item provide the evidence.
\end{enumerate}

\subsubsection{Protect the machine}
The highest priority is clearly to avoid any damage to accelerator equipment.

\subsubsection{Protect the beam}
The objective is to maximize beam time, but complex protection systems reduce the availability of the machine. The number of `false` interlocks stopping operation must be minimized. This is a trade-off between protection and operation. A `false` interlock is defined as an interlock that stops operation even though there is no risk (example: a temperature sensor reading a wrong value, therefore switching off the power converter of a magnet and stopping beam operation).

\subsubsection{Provide the evidence}
If the protection systems stop operation (e.g.\ dump the beam or inhibit injection), clear diagnostics should be provided \cite{Zerlauth2009}. If something goes wrong (leading to damage, but also a near miss), it should be possible to understand the event. This needs synchronized transient recording of all the important parameters in all relevant systems, as well as long-term logging of parameters with reduced frequency (such as 1~Hz). Examples are the current in all magnets, beam position, beam losses and beam intensity. The frequency of transient recording depends on the system and can go from Hz to MHz.

\subsection{Active and passive protection}
\subsubsection{Active protection}
Active protection starts with the detection of the failure by a sensor that is monitoring the parameters of equipment (for example, detecting a failure of a power converter) or by beam instrumentation detecting when the beam starts to be affected by the failure (for example, increased beam losses or a different orbit). The beam must be turned off as soon as possible with an actuator. This can be done in different ways, for example by switching off the source or the RF system. In the case of an accelerator complex with a chain of several accelerators, injection of the beam into the next stage of the accelerator complex should be prevented in case of failure. When the beam is stored in a synchrotron or storage ring, the beam must be extracted onto a target that can accept the beam pulse without being damaged (beam dump).

\subsubsection{Passive protection}
There are failures (e.g.\ ultra-fast losses) when active protection is not possible. One example is the protection against mis-firing of an injection or extraction kicker magnet. A beam absorber or collimator is required to stop the mis-kicked beam in order to avoid damage. All possible beam trajectories in such case must be considered, and the absorbers must be designed to absorb the beam energy without being damaged. Another example is a fast extraction of a high-intensity beam from a circular accelerator into a transfer line. When the extraction takes place, the parameters of the transfer line must be correctly set since in case of a wrong magnet current the beam could be deflected into the vacuum chamber.

The best strategy is to prevent a specific failure from happening. As an example, fast diagnostic kicker magnets that could deflect the beams into the vacuum chamber wall should only be installed in high-intensity machines if they are indispensable.

Failure should be detected as early as possible, with priority at the hardware level. For most failures, this allows stopping beam operation before the beam is affected. This requires monitoring of the hardware (such as state signals, parameters etc). As an example, a trip of a magnet power converter should be detected as early as possible.

It is not always possible to detect failures at the hardware level. The second method is to detect the initial consequences of a failure with beam instrumentation and to stop the beam before equipment is damaged. This requires reliable beam instrumentation.

When a failure is detected, beam operation must be stopped. For synchrotrons and storage rings the beam is extracted by a fast kicker magnet into a beam dump block. Injection must be stopped. For linacs the beam is stopped in the low-energy part of the accelerator by switching off the source, deflecting the low-energy beam by electrostatic plates (`choppers') or by switching off the RFQ for proton linacs.

An electronic system (beam interlock system)  links the different protection system. It ensures that the beam is extracted from a synchrotron, injection is stopped, RF acceleration might be stopped (for linacs). The interlock system might include complex logic.

\subsection{Design considerations for protection systems}

There are several principles that should be considered in the design of protection systems, although it might not be possible to follow all these principles in all cases.
\begin{enumerate}

  \item If the protection system does not work, it is better stopping operation rather than continuing and risking damaging equipment.

  \item Fail-safe design: in case of a failure in the protection system, protection functionalities should not be compromised. As an example, if the cable that triggers the extraction kicker of the beam dumping system is disconnected, operation must stop.

  \item Detection of internal faults: the protection system must monitor the internal status. In case of an internal fault, the fault should be reported. If the fault is critical, operation must be stopped.

  \item Remote testing should be an integral part of the design, for example between two runs. This allows verification of the correct status of the system.

  \item Critical equipment should be redundant (possibly diverse redundancy, with the same or similar functions executed by different systems).

  \item Critical processes for protection should not rely on complex software running under an operating system and requiring the general computer network.

  \item It should not be possible to remotely change the most critical parameters. If parameters need to be changed, the changes must be controlled and logged and password protection should ensure that only authorized personnel can do the change.

  \item Safety, availability and reliability of the systems should be demonstrated. This is possible by using established methods to analyze critical systems and to predict failure rates.

  \item Operate the protection systems early on before they become critical, to gain experience and to build up confidence. This could be done before beam operation, or during early beam operation when the beam intensity is low.

  \item It is inevitable to disable interlocks (e.g.\ during the early phase of commissioning and for specific tests). Managing interlocks (e.g.\ disabling) is common practice. Keep track and consider it in the system design. Example for LHC: masking of some interlocks is possible, but only for low-intensity/low-energy beams (`safe beams').
\end{enumerate}

\section{Beam instrumentation for machine protection}
\subsection{Beam loss monitors}
Beam loss monitors (BLMs) are used for monitoring beam losses to understand the performance of the accelerator as well as for machine protection. If used for protection, it is important that the monitors cover the entire accelerator and there is no region without BLMs where beam losses can occur.

\subsection{Beam position monitors}
BPMs ensure  that the beam has the correct orbit. In most cases the beam should be centred in the aperture. There are some exceptions, for example for extracting beams from an accelerator. The beam must have an offset, e.g.\ a closed orbit bump is applied to position the beam close to a septum magnet. This must work reliably, otherwise the extracted beam might hit equipment. BPMs monitor the amplitude of such bumps and are effectively redundant monitors to the sensors measuring the magnet current in the closed orbit dipoles.

\subsection{Beam current monitors}
If the beam transmission between two locations of the accelerator complex is too low (if the beam is lost somewhere between): stop beam operation. Differential BCMs are of particular importance for linear accelerators.  At low energy (below about 50~MeV/$c$) beam loss monitors are not applicable and machine protection relies on differential beam current measurements. If the beam lifetime in a synchrotron or storage ring is too short: dump the beam.

\subsection{Beam size monitors}
If the beam size is too small, this could be dangerous for windows, targets etc. The monitor ensures correct beam size. A too small beam size could also lead to an underestimation of the damage potential of the circulating beam.

\section{Machine protection for different types of accelerators}
High-energy hadron synchrotrons and colliders can have large stored energies in the beams. In case of a failure the energy stored in the beams must be safely discharged. In general, the beams are extracted onto a target that can withstand the full energy stored in the beam. Safely dumping the beams can be rather challenging. This is an issue for accelerators such as the LHC \cite{TheLHCStudyGroup1995}, HERA \cite{Werner2006}, Tevatron \cite{Mokhov2006}, RHIC \cite{Fedotov2007} and SPS \cite{Goddard2006}.

Linear colliders and other accelerators with very high beam power densities due to small beam size have other challenges. Examples are the SLAC linac \cite{Koontz2006}, International Linear Collider (ILC), CLIC (Compact Linear Collider), Next Linear Collider (a project that was discussed before ILC) \cite{Ross2000, Ross2000a} and FLASH \cite{Froehlich2006, Froehlich2006a} (average power of 50~kW).
For an accelerator like the ILC one beam pulse can lead  to damage. A citation from \cite{Ross2000a} discussing the use of a low-intensity pilot beam during operation says: ``for any time interval large enough to allow a substantial change in the beam trajectory of component alignment (fraction of a second), a pilot beam must be used to prove the integrity of the accelerator''.

In synchrotron light sources with high-intensity beams and secondary photon beam, the primary beam, but also the intense photon beam, can damage equipment.

For energy recovery linacs, we take as an example the prototype of such accelerators in Daresbury: a single bunch train cannot damage equipment, but in case of beam loss the next train must not leave the (injector) station \cite{Buckley2006}.

Owing to the low intensity of medical accelerators there is no risk of damage to equipment. However, it is vital to prevent a too high dose to patients. The strategies and techniques to minimize risks for humans are similar to those for machine protection.

For very short bunches with high current the beam induces image currents in surrounding equipment that can damage bellows, beam instruments, cavities etc.

%TIM

\subsection{Examples from LHC}
Machine protection starts with a careful commissioning of the magnet powering system, considering that an energy of about 10~GJ is stored in the superconducting magnets. Magnet protection and powering interlocks must be operational long before starting beam operation. The strategy for LHC machine protection when operating with beam reflects many of the principles that have been discussed is as follows.

\begin{enumerate}

  \item Definition of aperture by collimators.

  \item Early detection of failures of equipment acting on beams generates a beam dump request, possibly before the beam is affected.

  \item Active monitoring of the beam parameters with beam instruments detecting abnormal
  beam conditions and generating beam dump requests within a single machine turn.

  \item Reliable transmission of beam dump requests from a large variety of systems to the beam dumping system. An active signal is required for operation; the absence of the signal is considered as a beam dump request and injection inhibit.

  \item Reliable operation of the beam dumping system for dump
  requests or internal faults, safely extracting the beams onto external dump blocks.

  \item Passive protection by beam absorbers and collimators for specific cases of failure.

\end{enumerate}

\begin{figure}
\centering
\includegraphics[width=.78\linewidth]{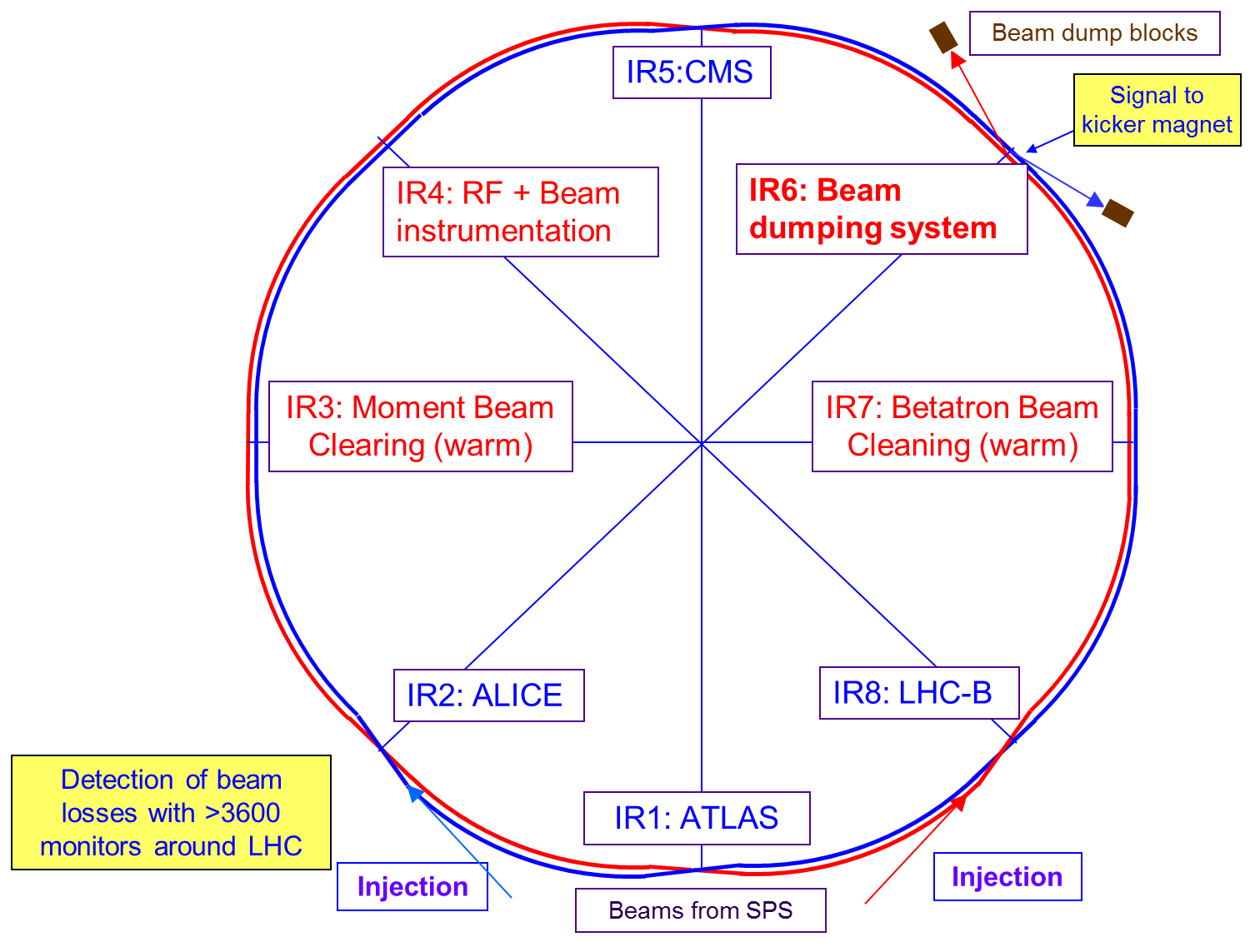}
\caption{Layout of LHC with some of the systems for machine protection}
\label{LHC-protection-overview}
\end{figure}

The general architecture for the essential elements in the machine protection system at LHC is shown in Fig. \ref{LHC-protection-overview}. The LHC has eight sectors and eight insertions; three sectors are related to machine protection: two cleaning insertions with a large number of collimators and one insertion for the beam dumping system. More than 3600 beam loss monitors are installed around the machine. In case of a failure detected by hardware or beam monitors, the beam interlock system transmit a beam dump request to the beam dumping system and the beams are extracted.

\subsubsection{Beam dumping system at LHC}
\begin{figure}
\centering
\includegraphics[width=.8\linewidth]{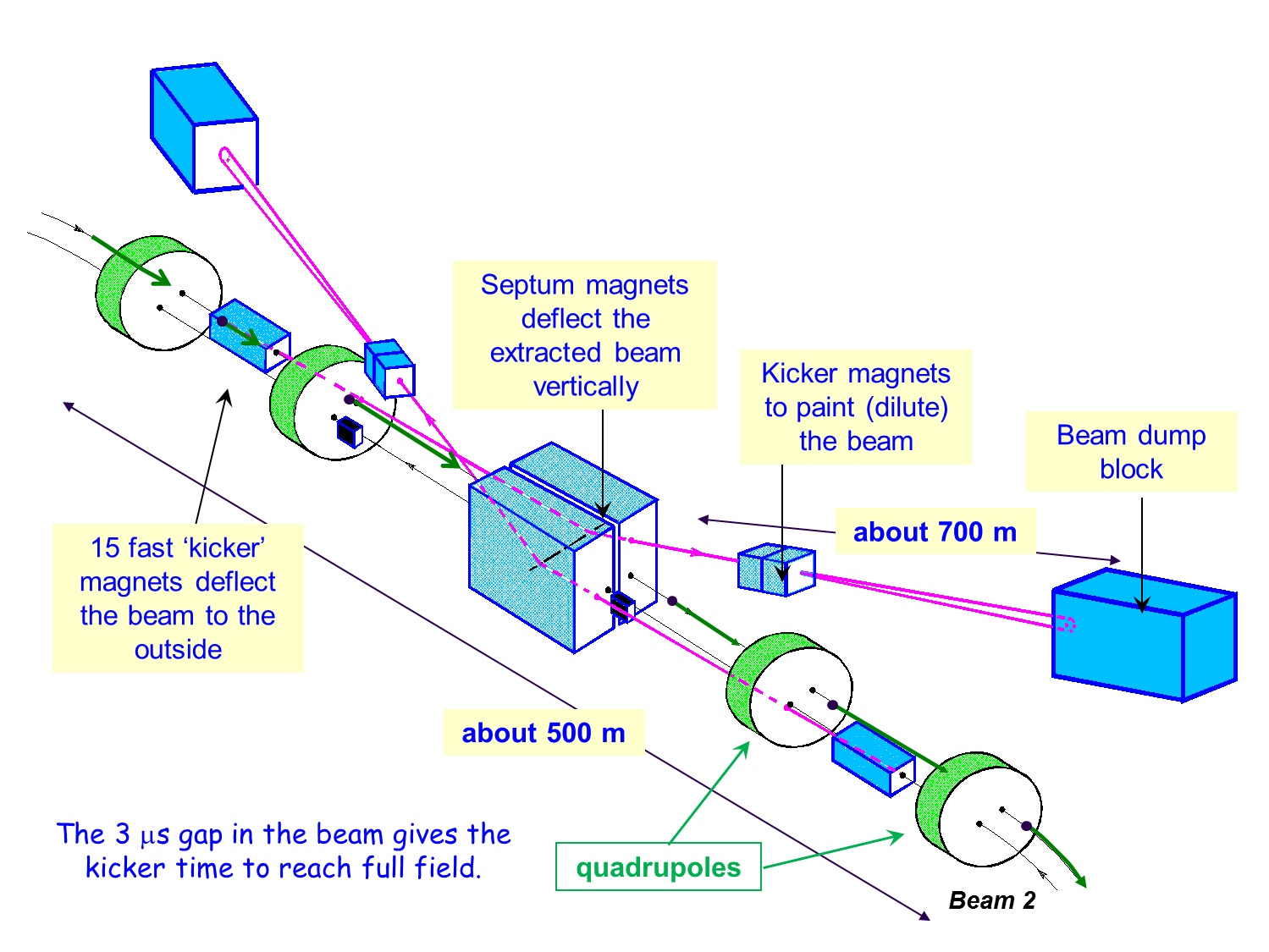}
\caption{Layout of the beam dumping systems, for both LHC beams (courtesy of M.~Gyr)}
\label{Beamdump-Layout}
\end{figure}

The role of the LHC beam dumping system \cite{Uyhoven2008} is to safely dispose of the beam when beam operation must be interrupted for any reason. Fifteen fast kicker magnets with a pulse rise time of less than 3~$\mu$s deflect the beam by an angle of 280~$\mu$rad in the horizontal plane, see Fig.~\ref{Beamdump-Layout}. To ensure that all particles are extracted from the LHC without losses, the beam has a particle-free abort gap with a length of 3~$\mu$s corresponding to the kicker rise time. The extraction kickers are triggered such that the field increases from zero to the nominal value during this gap when there should be no particles.

Downstream of the kickers the beam is deflected vertically by 2.4~mrad towards the beam dump block by 15 septum magnets. A short distance further downstream, 10 diluter kicker magnets are used to paint the bunches in both horizontal and vertical directions to reduce the beam density on the dump block. The beam is transferred through a 700~m long extraction line to increase the transverse r.m.s.\ beam size from approximately 0.2 to 1.5~mm and to spread the bunches further on the dump block.

The overall shape is produced by the deflection of the extraction and dilution kickers. For nominal beam parameters, the maximum temperature in the beam dump block is expected to be in the order of about 800$^{\circ}$C.

\subsubsection{Beam loss monitoring at LHC}
The monitors should be fast, for LHC down to 40 $\mu$s, in order to detect beam losses in time to stop operation. They should be designed such that they can trigger a beam dump and stop operation before very fast beam losses damage equipment. There are about 3600 chambers distributed over the ring to detect abnormal beam losses and if necessary trigger a beam abort \cite{Dehning2011}.

\begin{figure}
\centering
\includegraphics[width=.8\linewidth]{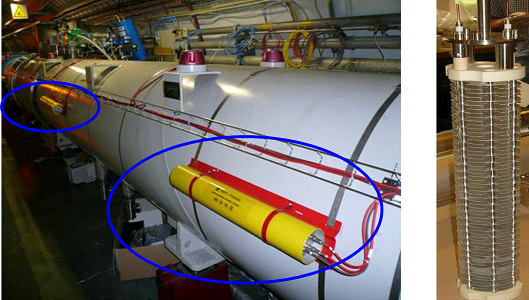}
\caption{Beam loss monitor at LHC}
\label{LHC-BLM}
\end{figure}

For LHC (Fig.~\ref{LHC-BLM}) and several other accelerators, ionization chambers are used to detect beam losses. The reaction time is down to microseconds;
they can have a very large dynamic range exceeding 10$^8$.

\begin{figure}
\centering
\includegraphics[width=.8\linewidth]{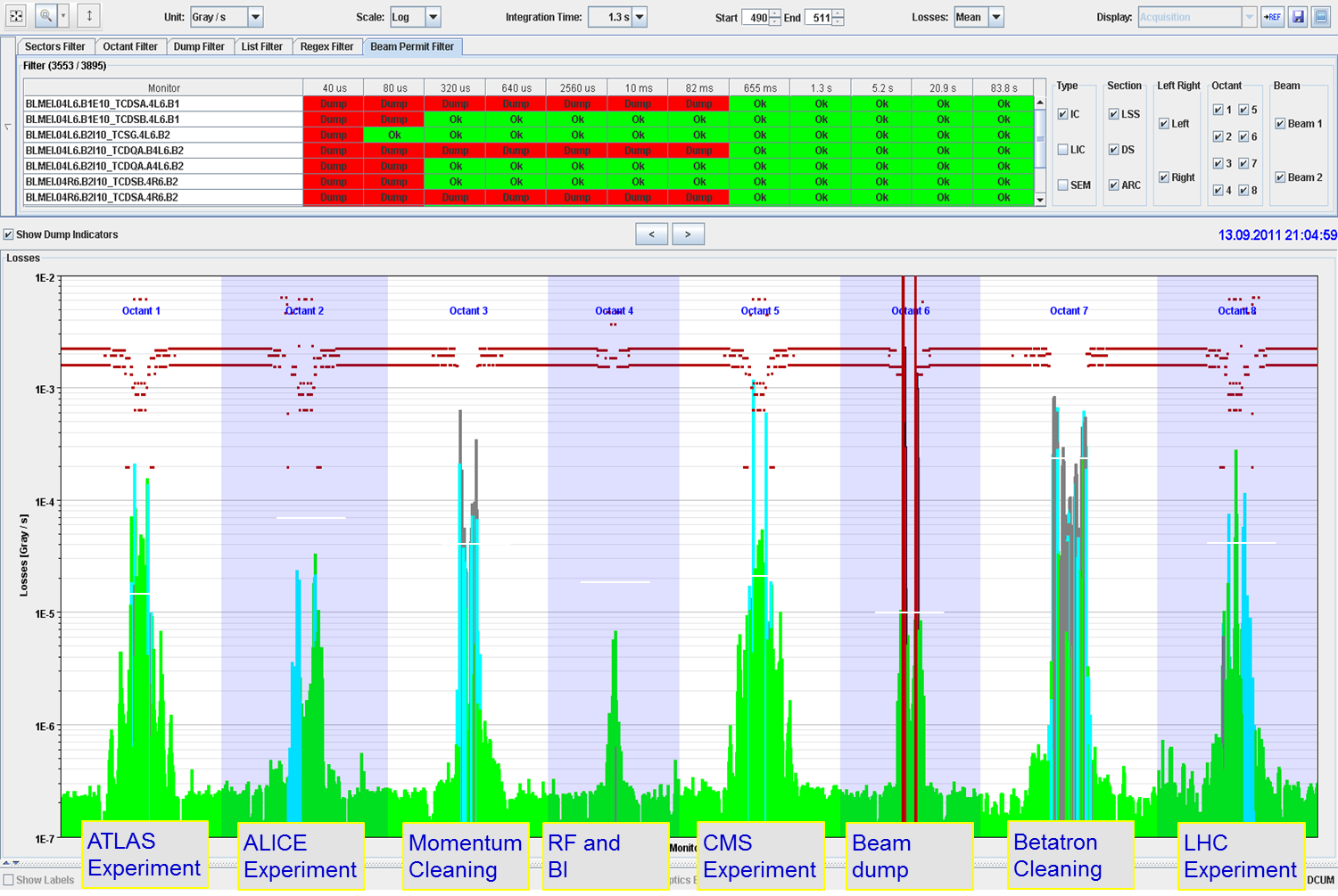}
\caption{Regular beam losses during luminosity operation}
\label{Beam-Losses-normal}
\end{figure}

\begin{figure}
\centering
\includegraphics[width=.8\linewidth]{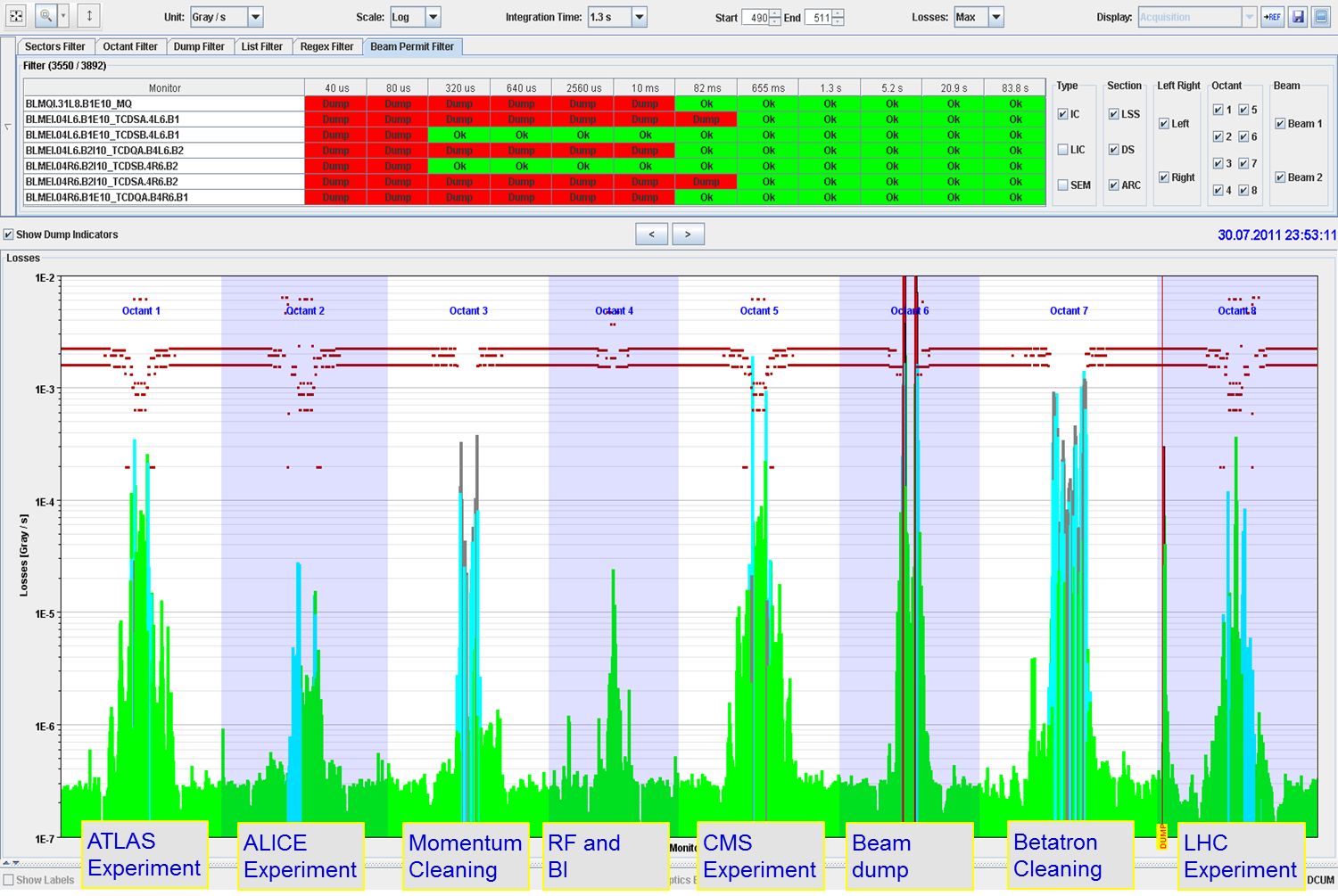}
\caption{Beam losses with UFO during luminosity operation}
\label{Beam-Losses-UFO}
\end{figure}

\begin{figure}
\centering
\includegraphics[width=.8\linewidth]{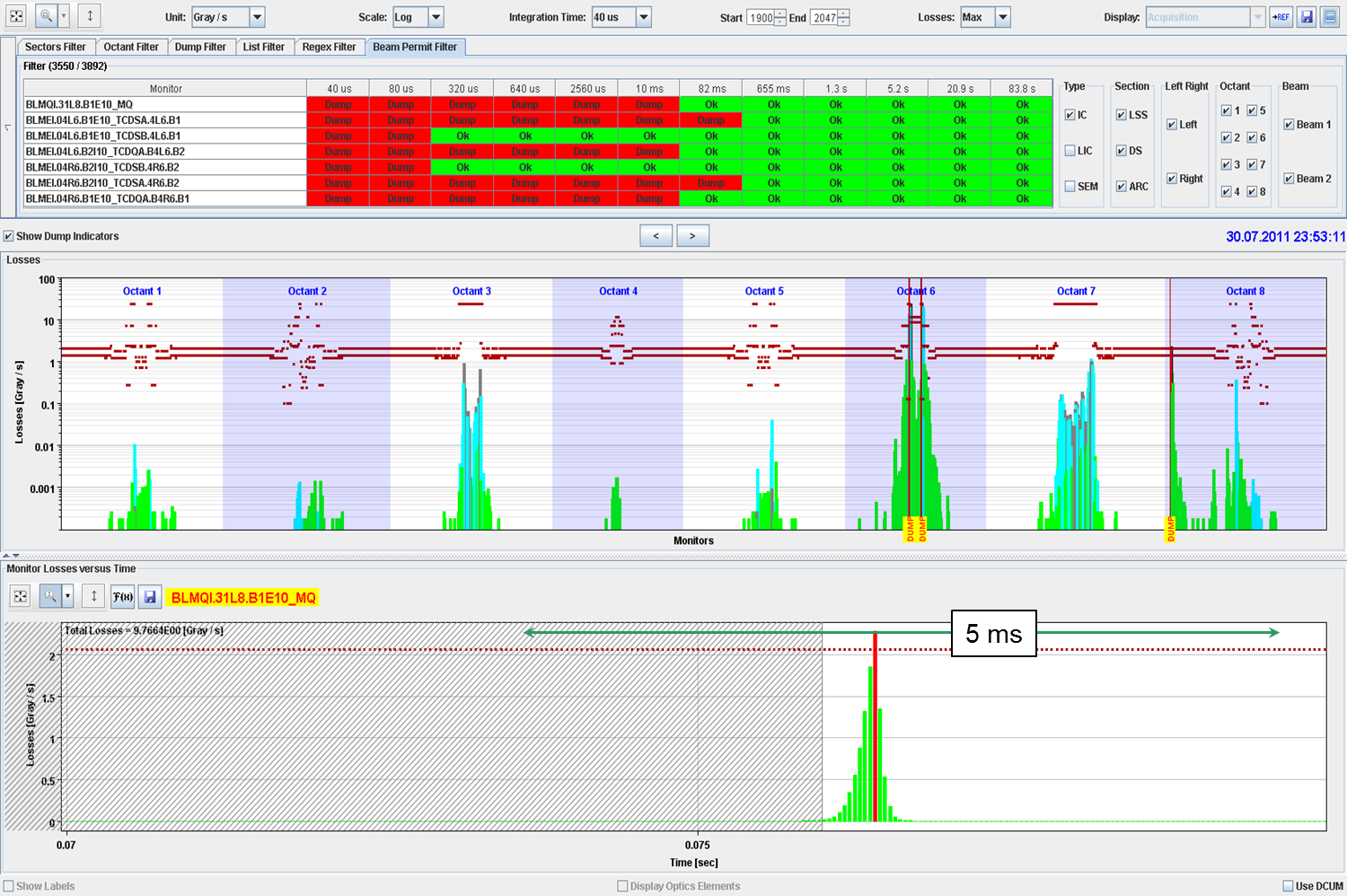}
\caption{Time structure of the beam losses caused by a UFO}
\label{UFO}
\end{figure}

Figure~\ref{Beam-Losses-normal} shows the beam losses around LHC during regular luminosity operation. Losses are low in the arcs with the superconducting magnets, higher in the insertions with the experiments due to the debris of the collisions and very high in the betatron cleaning insertions. Figure~\ref{Beam-Losses-UFO} shows the losses recorded after a beam dump during regular operation. A UFO in the arc between the betatron cleaning insertion and the LHCb experiment caused locally an increase of beam losses. The UFO (dust particles getting into the beam, see \cite{baer2013}) causes very fast losses and when the threshold of one BLM is exceeded, the beams are dumped (see Fig. \ref{UFO}).

\subsection{Protection for high power proton linacs}
High-power proton accelerators (e.g.\ spallation sources) operate with beam power of some 10~kW to above 1~MW. There is risk of both damage and activation. Examples are SNS \cite{Sibley2005}, PSI cyclotron \cite{Schmelzbach2006}, ISIS \cite{Findlay2006}, J-PARC \cite{Yoshikawa2007}), FRIB \cite{Wei2012} and ESS designed for 5~MW \cite{Peggs2005}.

In case of failure the beam must be switched off. The time needed to switch off the beam before equipment could be damaged depends on particle momentum, beam power, the type of failure and the equipment exposed to beam losses.

\begin{figure}
\centering
\includegraphics[width=.8\linewidth]{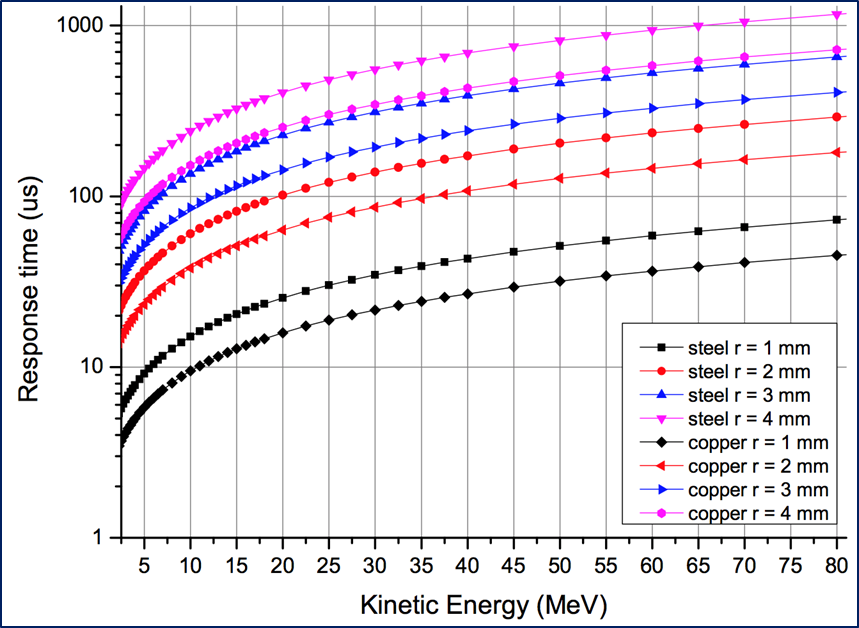}
\caption{Time to melt copper and steel, as a function of proton momentum for different beam sizes \cite{Tchelidze2012}}
\label{Laliplot}
\end{figure}

In case of an uncontrolled beam loss during 1~ms at ESS the deposited energy is up to 130~kJ, for 1~s up to 5~MJ. It is required to inhibit the beam after detecting uncontrolled beam loss as fast as possible. There is some delay between detection of a failure (e.g.\ detection of beam losses by a BLM) and `beam off'. Figure \ref{Laliplot} shows the time to melting of steel and copper in the case where the proton beam hits a metal surface between 3 and 80~MeV/$c$ \cite{Tchelidze2012}.

For example, after the DTL normal-conducting linac, the proton energy is 78~MeV/$c$. In case of a beam size of 2~mm radius, melting would start after a beam impact of about 200~$\mu$s. Inhibiting of the beam after a failure is detected should be in about 10\% of this time.

\begin{figure}
\centering
\includegraphics[width=.8\linewidth]{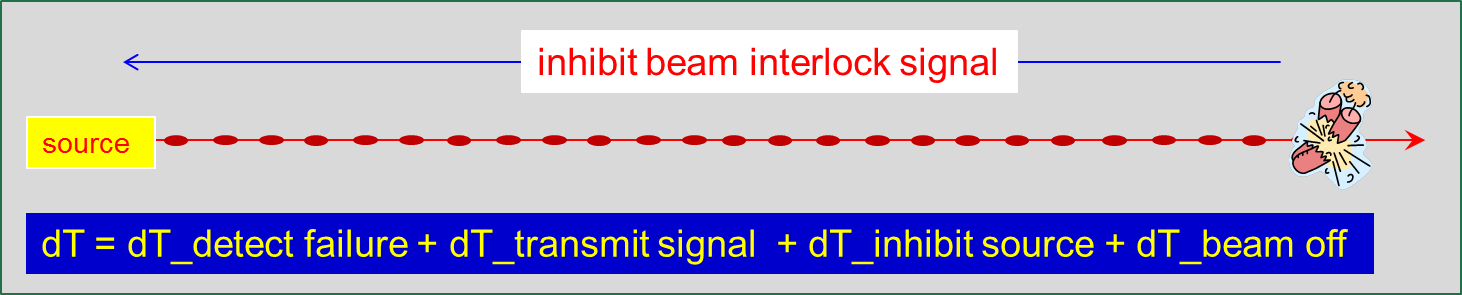}
\caption{Time to stop the beam in a linac}
\label{Time-stop-beam}
\end{figure}

The time between detecting a failure and stopping the beam has to be considered. Assume that the beam is lost and the losses are detected by BLMs at the end of the linac. It takes some time to detect the losses and to generate an interlock signal ${\rm d}T_{\rm failure~detection}$, to propagate the interlock signal to the actuator that stops the beam ${\rm d}T_{\rm signal~transmission}$ and to stop the beam ${\rm d}T_{\rm inhibit~beam}$. There will be a number of bunches between source and end of the linac that cannot be stopped (${\rm d}T_{\rm beam~off}$). The total time that the beam will still be lost is given by (see also Fig. \ref{Time-stop-beam})
\begin{equation}
  {\rm d}T = {\rm d}T_{\rm failure~detection} + {\rm d}T_{\rm signal~transmission}  + {\rm d}T_{\rm inhibit~beam} + {\rm d}
  T_{\rm beam~off}.
\end{equation}

Depending on the length of the linac, this can be very long and might require intermediate beam dump systems in case of accelerators such as ILC.

\section{Conclusions}
There has been a significant increase in the number of papers on machine protection during particle accelerator conferences,
from a few papers per conference in the 1990s to more than 100 papers in recent conferences.

Machine protection goes far beyond the equipment protection and across many systems. It requires the understanding of many different types of failures that could lead to beam loss. It requires fairly comprehensive understanding of all aspects of the accelerator (accelerator physics, operation, equipment, instrumentation) and touches many aspects of accelerator construction and operation.

Machine protection is becoming increasingly important for future projects, with increased beam power and energy density (W/mm$^2$ or J/mm$^2$) and increasingly complex machines.

Protection of equipment, even when operating without beam, must not be forgotten. The largest accident happening at an accelerator was the rupture of a superconducting cable at LHC in 2008 due to the very large energy stored in the superconducting magnet system.

\section{Acknowledgements}
I wish to thank many colleagues from CERN and ESS and the authors of the listed papers for their help and for providing material for this article.
 This paper is an updated version of the lecture given at a CAS in 2008 \cite{Schmidt2008}.

%\section{Bibliography}

%\bibliographystyle{ieeetr}
%\bibliography{\myreferences/Rudi-bibliography,\myreferences/Tahir,\myreferences/SPS-bibliography,\myreferences/Other-papers}

\end{document}